\documentclass{ws-procs9x6}
\begin{document}

\begin{center}
To appear on the Proceedings of the 13th ICATPP Conference on\\
Astroparticle, Particle, Space Physics and Detectors\\ for Physics Applications,\\ Villa  Olmo (Como, Italy), 3--7 October, 2011, \\to be published by World Scientific (Singapore).
\end{center}
\vspace{-1.5cm}

\title{HELIOSPHERE DIMENSION AND COSMIC RAY MODULATION}

\author{P. Bobik$^{1}$, M.J. Boschini$^{2,4}$, C. Consolandi$^{2}$,
S. Della Torre$^{2,5}$, M. Gervasi$^{2,3,*}$, D. Grandi$^{2}$, K.
Kudela$^{1}$, F. Noventa$^{2,3}$, S. Pensotti$^{2,3}$, P.G.
Rancoita$^{2}$, D. Rozza$^{2,3}$}

\address{$^{1}$ Institute of Experimental Physics, Kosice (Slovak Republic) \\
$^{2}$Istituto Nazionale di Fisica Nucleare, INFN Milano-Bicocca, Milano (Italy) \\
$^{3}$Department of Physics, University of Milano Bicocca, Milano (Italy) \\
$^{4}$CILEA, Segrate (MI) (Italy) \\
$^{5}$Department of Physics and Maths, University of Insubria, Como (Italy)\\
*E-mail: massimo.gervasi@mib.infn.it}

\begin{abstract}
The differential intensities of Cosmic Rays at Earth were calculated
using a 2D stochastic Montecarlo diffusion code and compared with
observation data. We evaluated the effect of stretched and
compressed heliospheres on the Cosmic Ray intensities at the Earth.
This was studied introducing a dependence of the diffusion parameter
on the heliospherical size. Then, we found that the optimum value of
the heliospherical radius better accounting for experimental data. We
also found that the obtained values depends on solar activity. Our
results are compatible with Voyager observations and with models of
heliospherical size modulation.
\end{abstract}


\bodymatter
\section{The 2D Model of the Heliosphere}
HelMod Code\cite{HelMod2011} solves the bi-dimensional Parker's
particle transport equation\cite{parker1965}\!. A Monte Carlo
technique is applied on a set of Stochastic Differential Equations
(SDEs) fully equivalent to the Parker's
equation\cite{AstraArticle2011}\!. The model takes into account
particle drift effects and latitudinal dependence of the solar
wind speed and of the Interplanetary Magnetic Field (IMF). It is
described in details in Ref.~\refcite{HelMod2011}. In the model,
the IMF from Parker\cite{parker1958} is modified introducing a
small latitudinal components as described in
Ref.~\refcite{langner2004}. For periods of low solar activity, we
take a solar wind speed gradually increasing from the Earth
position up to a maximum value near the heliospherical poles
($\simeq 760$\,km/s)~\cite{McComas2000}\!. For periods approaching
the solar maximum we assume a solar wind speed independent on the
latitude.

The symmetric part of the diffusion tensor, in a reference frame
with one axis aligned with the Parker's magnetic-field, is purely
diagonal containing transverse ($K_{\perp \theta}$ and $K_{\perp
r}$) and parallel ($K_{||}$) components~\cite{potgieter2000}\!. The
diffusion coefficients are given by~\cite{potgieter1994}

\begin{eqnarray}\label{duffusion_comp}
\begin{array}{l}
K_{||} = \beta \, K_0(t) \,K_{P}(P)\,\left[\frac{B_{\oplus}}{3B}
\right] \ , \\
K_{\perp r} = \rho_k \ K_{||} \ , \\
K_{\perp \theta} = \iota(\theta) \ \rho_k \ K_{||} \ ,
\end{array}
\end{eqnarray}

\noindent where $\beta = v/c$, $v$ the particle velocity and $c$ the
speed of light; the diffusion parameter $K_0$ accounts for the
dependence on the solar activity; $ B_{\oplus}$ is the measured
value of IMF at the Earth position - typically $ \approx 5$\,nT, but
changing with time - obtained from Ref.~\refcite{SW_web}; $B$ is the
magnitude of the large scale IMF as a function of heliocentric
coordinates; finally, the term $K_{P}$ takes into account the
dependence on the rigidity $P$ of the GCR particle usually expressed
in GV.~In the present model $K_P \approx P$ (e.g., see Ref.\cite{perko1987}).
Furthermore \ $\rho_k = 0.05$ \ and, as described in
Ref.~\refcite{HelMod2011},

\begin{eqnarray}\label{enhaced_K}
\iota(\theta)  = \left\{
\begin{array}{rll}
10 \ ,  & \textrm{ in the polar regions},\\
1 \ , & \textrm{ in the equatorial region}.
\end{array}\right.
\end{eqnarray}

After the transformations from 3D field-aligned into 2D heliospherical
coordinates~\cite{burg2008}\!, the symmetric components of the
diffusion tensor contains both diagonal ($K_{r r}$ and $K_{\theta
\theta}$) and off-diagonal terms ($K_{r \theta}$ and $K_{\theta
r}$), resulting by a proper combination of $K_{\perp \theta}$,
$K_{\perp r}$ and $K_{||}$~\cite{HelMod2011}\!.

\section{The Diffusion Parameter}
$K_0$ accounts for the dependence on the solar activity. We
estimated $K_0$ by using the modulation strength $\phi_{\rm s}$, in
the framework of the Force Field (FF)
approximation~\cite{gleeson1967}\!. $\phi_{\rm s}$ was evaluated
starting from Neutron Monitor (NM) counting rates in
Ref.~\refcite{ilya05}. We moreover used a practical correlation of
$K_0$ with the level of solar activity in the different solar
phases~\cite{HelMod2011}\!. We used, as solar activity monitor, the
Smoothed Sunspot Number (SSN).

This method is sensitive to the modulation of the GCR flux
integrated over the full heliosphere, from the outer boundary to the
Earth position, down to a lower limit in rigidity of $\sim$ (2-3)
GV. This limit is fixed by the sensitivity of the NM network, due to
the geomagnetic rigidity cut-off and to the atmospheric yield
function. The outer boundary of the heliosphere is located at the
position of the Termination Shock. Beyond this limit the model of
heliosphere we are using is not more valid. Moreover, the additional
modulation occurring in the heliosheat only affects particles with
rigidity well below 1 GV~\cite{Bobik2008,Sherer2011}\!. The method
is sensitive also to the LIS used for the estimation of the
modulation strength, but several LIS spectra do not differ each
other above this rigidity limit. Finally the diffusion parameter
depends on the outer boundary position, as follows from the FF
approximation:

\begin{equation}\label{md_par_red2}
K_0(t) = \frac{V_{\rm sw}(t) \left(R_{\rm TS}-R_{\rm earth}
\right)}{3 \phi_{\rm s}(t)} \ .
\end{equation}

In Ref.~\refcite{HelMod2011} the boundary of heliosphere was placed
at 100\,AU. The solar cavity was split in 15 spherical regions to
take into account the time spent by SW to travel outward. In each
region of the interplanetary space, the parameters (i.e., SW speed,
SSN, $B_\oplus$, tilt angle) are related to the solar activity at
the time of the injection of the solar wind diffusing in that
region~\cite{DavArticle2010}\!. In this way modulated intensities of
protons, down to $\sim 400$ MeV, were simulated and successfully
compared with experimental data covering roughly one solar cycle. We
did not find significant differences changing the position of the
outer boundary of the heliosphere~\cite{HelMod2011}\!.

\section{Heliospherical Size and Diffusion Parameter}
In the past years the position of the Termination Shock was
estimated through the observations of Voyager 1 and Voyager 2
spacecrafts (see Refs.~\refcite{Voyager1,Voyager2}
and Table~\ref{aba:tbl1}). In addition several authors (see
Refs.~\refcite{WB2000,WB2004}) suggest that the
size of the heliospere should change with the solar activity,
following a quasi-periodic feature, roughly anti-correlated with the
SSN.

\begin{table}
\tbl{Voyager crossings of Termination Shock.}
{\begin{tabular}{ccc} \toprule
 & $R_{TS}$ (AU) & solar latitude (deg) \\
\colrule
Voyager 1 & 94.0 &  + 34.3 \\
Voyager 2 & 83.7 &  $-$ 27.5 \\
\botrule
\end{tabular}
} \label{aba:tbl1}
\end{table}

\noindent Following these results we evaluated the effect of
stretched and compressed heliospheres on the Cosmic Ray intensities
at the Earth introducing a dependence of the diffusion parameter on
the heliospherical size. We defined a new diffusion parameter $K_0^*$,
introducing the parameter $r(R_{TS},P)$ sensitive to the position of
the Termination Shock:

\begin{equation}\label{ratio1}
K_0^*(R_{TS}) = r(R_{TS},P) \ K_0(100 \ \textrm{AU})
\end{equation}

\begin{equation}\label{ratio2}
r(R_{TS},P) = 1 + f(P)\left[\frac{R_{TS} (\textrm{AU}) - 100}{99} \right]\ .
\end{equation}

\noindent $r(R_{TS},P)$ allows to modify the value of the diffusion
parameter adapting it to a different volume of the heliosphere,
determined by $R_{TS}$. $r(R_{TS},P)$ is fully effective below a
rigidity limit $P_{1}$. We also defined a transition function
$f(P)$:

\begin{eqnarray}\label{ratio3}
f(P) = \left\{
\begin{array}{lll}
0 \ , & \ \ \ \ \ \ \ \ \ \ \ \ & \textrm{for} \ \ \ P \geq P_{2} \ , \\
%& & \\
(P_{2} - P) / (P_{2} - P_{1}) \ , &  & \textrm{for} \ \ \ P_{1} < P < P_{2} \ , \\
%& & \\
1 \ , & & \textrm{for} \ \ \ P \leq P_{1} \ .
\end{array}\right.
\end{eqnarray}

\noindent For rigidity higher than $P_{2}$, the dependence on
$R_{TS}$ can be neglected. Here the diffusion parameter is still
defined for an heliospherical dimension of 100 AU. The dependence on
the heliospherical radius $R_{TS}$ is then effective at rigidity lower
than $P_{2}$. Using the novel diffusion parameter $K_0^*(R_{TS},P)$
we simulated the modulated spectra, for different values of
$R_{TS}$, $P_{1}$ and $P_{2}$, extending the modulated spectra down
to a lower rigidity.

\section{Results}
\begin{figure}
\centering
%\begin{sidewaysfigure}
%\scalebox{0.5}{
  \psfig{file=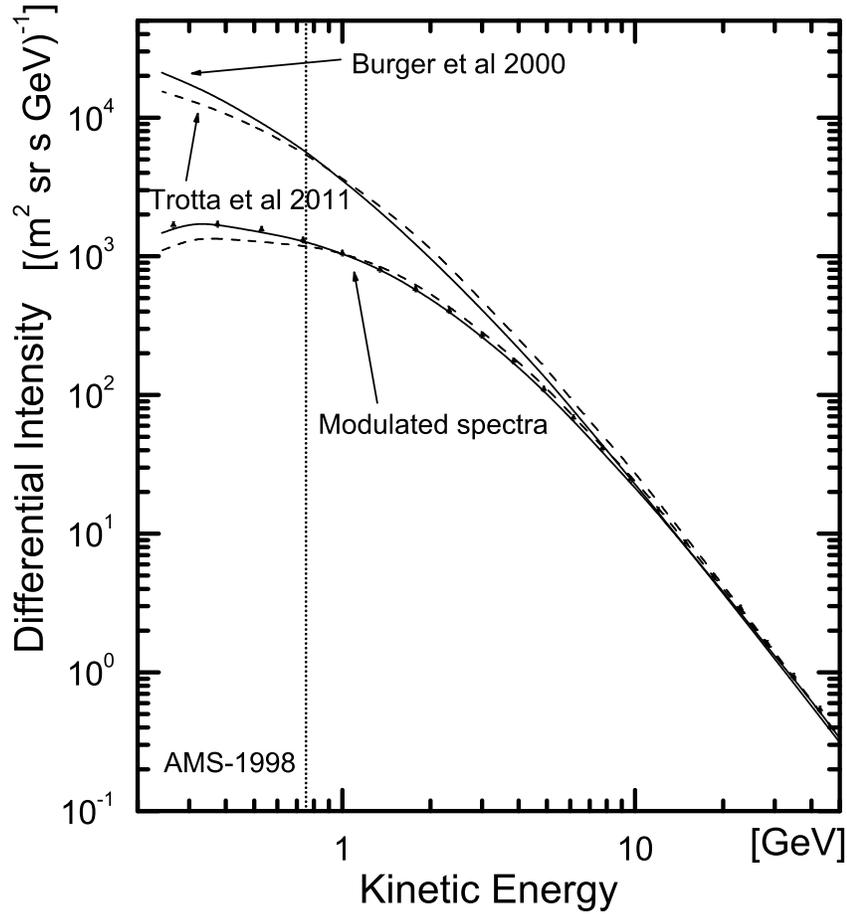,width=1.0\textwidth}
 \caption{Modulated proton spectra and comparison with AMS-01 data~\cite{AMS01_prot}\!.
 LIS are taken form Ref.~\refcite{Burger2000} and GALPROP~\cite{Galprop2011}\!.
 The vertical dotted line represents the lower limit of the sensitivity of the NM.
 Above this limit the two LIS are not significantly different. }
 \label{fig:1}
%\end{sidewaysfigure}
\end{figure}

We compare our simulated spectra with proton data extended down to a
kinetic energy of 200 MeV. Here we present results obtained using
the following rigidity parameters: $P_{2}=P_{1}=1.0$ GV. We used the
Local Interstellar Spectrum (LIS) from Ref.~\refcite{Burger2000} and
compared it with the LIS form GALPROP~\cite{Galprop2011}\!. In
Fig.~\ref{fig:1} the results compared with AMS-01
data~\cite{AMS01_prot} are shown, assuming $R_{TS} = 120$ AU, as
discussed later on.

We estimated the best value of $R_{TS}$, looking at the RMS
differences ($\eta_{\rm RMS}$) with experimental data:

\begin{equation}\label{eta_rms}
 \eta_{\rm RMS}=\sqrt{\frac{\sum_i \left(\eta_i/\sigma_{\eta,i}\right)^2}{\sum_i 1/\sigma^2_{\eta,i}}}\ ,
\end{equation}
with
\begin{equation}\label{eta_i}
 \eta_i=\frac{ f_{\rm sim}(T_i) -f_{\rm exp}(T_i) }{f_{\rm exp}(T_i)} \ ,
\end{equation}

\noindent where $T_i$ is the average energy of the $i$-th energy bin
of the differential intensity distribution and $\sigma_{\eta,i}$ are
the error bars including the experimental and Monte Carlo
uncertainties. For each experimental spectrum we got the best values
of $R_{TS}$ shown in Table \ref{aba:tbl2} together with the minimum
value of $\eta_{\rm RMS}$. Data from BESS flights are given in Ref.
\refcite{BESS}, data from AMS-01 are given in Ref.
\refcite{AMS01_prot}.

\begin{figure}
\centering
%\begin{sidewaysfigure}
%\scalebox{0.5}{
  \psfig{file=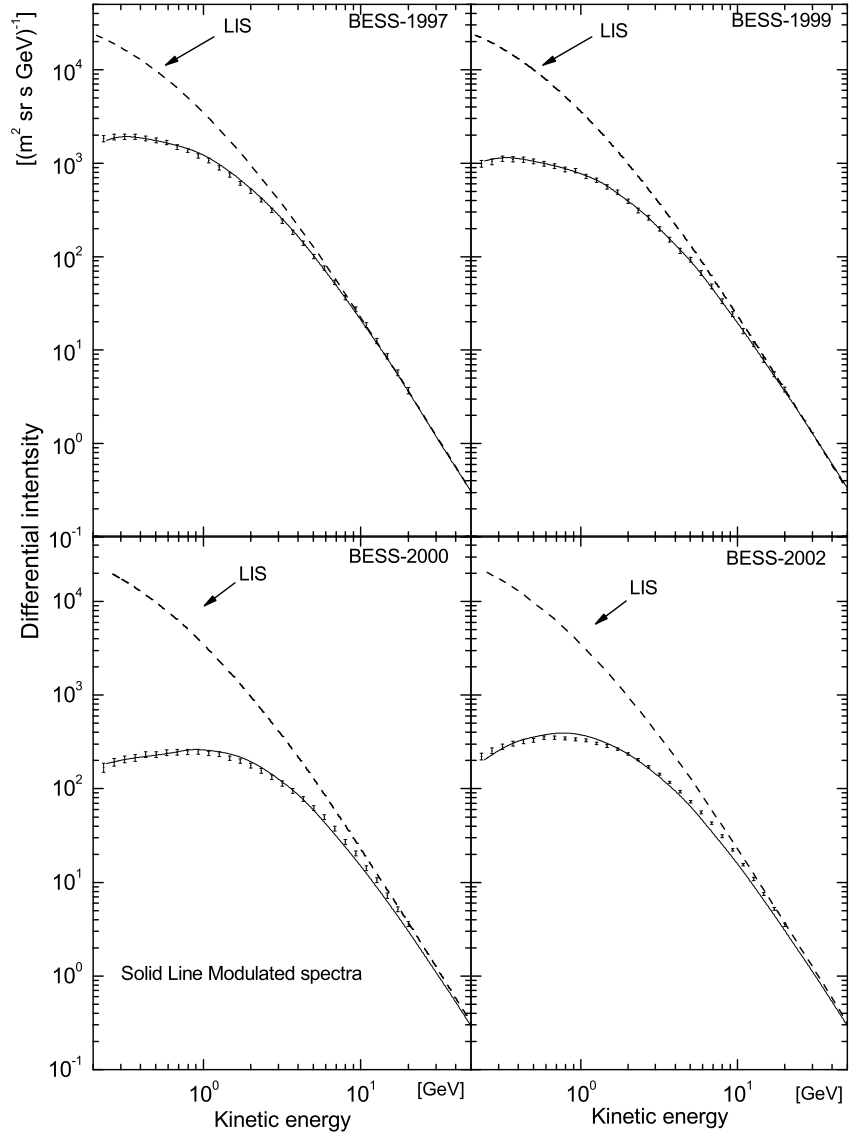,width=1.0\textwidth}
 % fig01.eps: 0x0 pixel, 300dpi, 0.00x0.00 cm, bb=
 \caption{Modulated proton spectra and comparison with BESS-1997,
 BESS-1999, BESS-2000, BESS-2002 observing data~\cite{BESS}\!.}
 \label{fig:2}
%\end{sidewaysfigure}
\end{figure}

\begin{table}
\tbl{Best values of $R_{TS}$, its minimum and maximum values, and
RMS differences between simulations and experimental data.}
{\begin{tabular}{ccccc} \toprule
 & $R_{TS}^{best}$ (AU) & $R_{TS}^{min}$ (AU) & $R_{TS}^{max}$ (AU) & $\eta_{\rm RMS}$ (\%) \\
\colrule
BESS-1997   & 115 & 100 & 130 &  7.05 \\
AMS-1998    & 120 & 110 & 135 &  4.86 \\
BESS-1999   & 120 & 110 & 130 &  3.35 \\
BESS-2000   & 140 & 125 & 150 & 10.00 \\
BESS-2002   & 105 &  95 & 115 & 11.78 \\
\botrule
\end{tabular}
} \label{aba:tbl2}
\end{table}

\begin{figure}
\centering
%\begin{sidewaysfigure}
%\scalebox{0.5}{
  \psfig{file=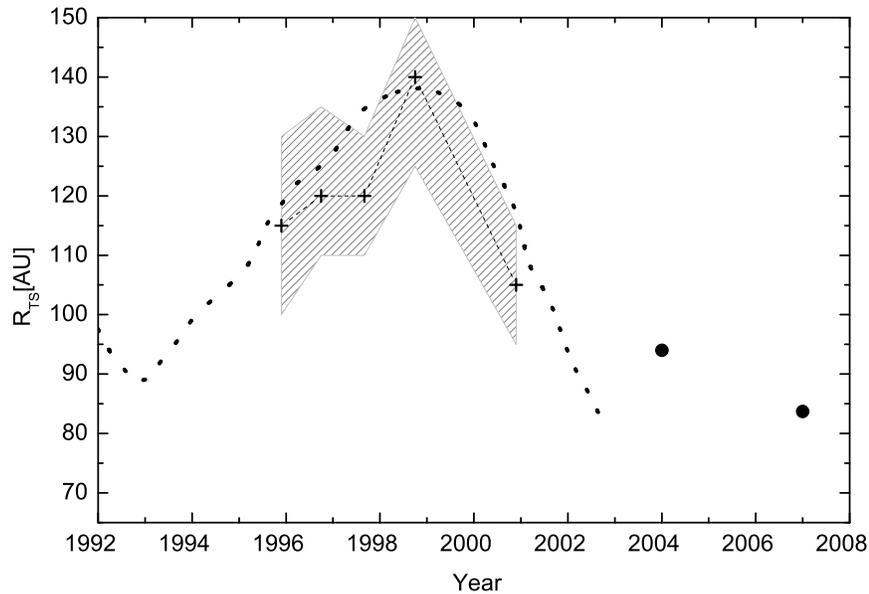,width=1.0\textwidth}
 \caption{$R_{TS}$ best value for the several experiments (crosses) in comparison
 with Voyager data~\cite{Voyager1,Voyager2} (dots) and models~\cite{WB2004}
 (dashed line). The shadow represents the region between the minimum and
 maximum value, as reported in Table \ref{aba:tbl2}.}
 \label{fig:3}
%\end{sidewaysfigure}
\end{figure}

In Fig.~\ref{fig:2} modulated spectra, obtained using values of
$R_{TS}$ reported in Table \ref{aba:tbl2}, are shown in comparison
with BESS experimental data. Modulated spectrum compared with AMS-01
data has been shown in Fig.~\ref{fig:1}. We did not use data
measured by Pamela~\cite{Pamela_Prot} because published spectra
start from 400 MeV, while our analysis is more sensitive below this
limit. In Table \ref{aba:tbl2} we report the interval of values of
$R_{TS}$ where $\eta_{\rm RMS}$ does not change by more than $\sim
(2-3)$ \% from its minimum value, reported in the last column. This
variation roughly represents the uncertainty of the computation
itself, and it is determined comparing simulations and data at
energies larger than $(10-20)$ GeV, i.e. above the region of solar
modulation. Results are shown in Fig.~\ref{fig:3} in comparison with
models~\cite{WB2004} and Voyager
measurements~\cite{Voyager1,Voyager2}\!.

As shown in Fig.~\ref{fig:1} and Fig.~\ref{fig:2}, modulated spectra
succeed to fit observing data, in particular during periods of low
solar activity. For more accurate results we need more accurate
experimental data. Current error bars are of the order of 5\% or
even larger. Moreover systematic deviations are present looking at
different data sets at energy above (20--30)\,GV, where spectra are not
affected by solar modulation. In addition current observation data,
taken on board of stratospheric balloons or space orbiters, may be
contaminated at low energy by secondaries produced inside the Earth
magnetosphere. A LIS spectrum with a slightly different shape could
be also preferred to fit better low energy data. Finally a
refinement of our model could be requested, starting with a slightly
different values of $P_{2}$ and $P_{1}$, in order to smooth the
ripple present in some spectrum. In the future a model with an
aspheric Heliosphere can be also developed.

\section{Conclusions}
We presented the HelMod 2D Monte Carlo code for the study of Cosmic
Rays propagation in the inner heliosphere. Both heliospherical shape
and size are supposed to be relevant for the modulation process. We
introduced a dependence of the diffusion parameter on the
heliospherical size, which accounts for the variation with time and
solar activity. We compare modulated spectra with experimental data
covering the solar cycle 23. Then we found, for our 2D model, the
best value of the heliospherical radius, which changes with time. Most
of the solar modulation occurs in the inner heliosphere and
differences in the heliospherical radius are effective only at energy
below a few hundred MeV. Our results are not in contradiction with
Voyager observations and models of TS distance as a function of
solar activity. We found that LIS form Ref.~\refcite{Burger2000}
fits better observation data at low energy.

\section*{Acknowledgments}
Authors acknowledge the use of NASA/GSFCs Space Physics Data
Facilitys OMNIWeb service, and OMNI data.

\bibliographystyle{ws-procs9x6}
\bibliography{DellaTorre2011_bib}

\end{document}